\documentstyle[multicol,aps,epsf]{revtex}
\newcommand{\ave}[1]{\langle #1 \rangle}

\newcommand{\sigsig}[2]{\sigma_{#1}\sigma_{#2}}

\newcommand{\itdelta}{{\it \Delta}}
\begin{document}
\title{A cluster heat bath method on a quasi-one dimensional Ising model}
\author{Osamu Koseki and Fumitaka Matsubara}
\address{Department of Applied Physics, Tohoku University, Sendai, 980-77}
\date{\today}
\maketitle
\begin{abstract}
We have proposed a cluster heat bath(CHB) method of a quasi-one dimensional Ising model and demonstrated that it reproduces the exact results of the two dimensional Ising model with axially anisotropic exchange interactions. 
The point of the method is to select one of the equilibrium spin 
configurations of each of the chains which are subjected by effective 
fields of surrounding spins. 
Thus the chains are always in their equilibrium state during the simulation 
and the $d$ dimensional model effectively becomes as a $d-1$ dimensional 
model of fictitious spins with their freedom of $2^N$ and the simulation time 
is drastically reduced, where $N$ is the number of the spins on the chain.  
\end{abstract}

\sloppy
\begin{multicols}{2}
\narrowtext

Recently, progress of the computational ability enable us to execute large scale computer simulations and more accurate nature of systems have been revealed.\cite{Binder}  However, low temperature phases of complex systems such as spin glasses have not well recognized yet, because they have many local minima and huge CPU time is necessary to perform computer simulations. To remove the difficulty, various algorithms have been proposed such as cluster algorithms\cite{Swendsen,Wolff} and extended ensemble methods.\cite{Berg,Marinari,Hukushima} 
Similar problems also occur in studying the phase transition even in {\it non-complex systems} whose ground state is known.  Among them, quasi-one dimensional magnetic systems are of great interest, because many attractive substances have been found such as CsCoCl$_3$,\cite{Mekata}  CsCoBr$_3$\cite{Yelon} and CsNiCl$_3$\cite{Johnson} for which successive phase transitions are observed.  
However, so long as we know, no effective algorithm has been proposed yet to treat those systems. 

In this letter, we discuss the difficulty faced on Monte Carlo(MC) 
simulations of qausi-one dimensional systems and propose a new effective method to remove it.  Our method is a cluster algorithm but different from previous ones\cite{Swendsen,Wolff} in which all the spins in the cluster are flipped simultaneously in accordance with the detailed valance condition.  In our algorithm, in contrast with this, one of the possible spin configurations of the cluster is selected according with its Boltzmann weight.  So we may call it as a cluster heat bath(CHB) method.
In the qausi-one dimensional system, this can be readily done, because the clusters may by chosen as one-dimensional chains and a recursion relation may be used to determine the spin configuration. 

We start with a two dimensional ferromagnetic Ising model with axially anisotropic nearest neighbor interactions  described by the Hamiltonian:
\begin{eqnarray}
    {\cal H} = - J  \sum_{i,j} S_{ij}S_{i+1j} 
               - J' \sum_{i,j} S_{ij}S_{ij+1},
\end{eqnarray}
where $J$ and $J'(\ll J)$ are coupling constants along the longitudinal and the transverse directions, respectively, and $S_{ij} = \pm 1$. 
This model was already solved by Onsager\cite{Onsager} about 50 years ago. 
However, the MC simulation of this model is not so easy. Suppose that we make a MC simulation of the model on an $M \times N$ lattice. As the temperature is decreased from a high temperature, the spin correlation in every chain along the 
longitudinal direction develops first. 
The correlation length $\xi(T)$ of the chain may be given as 
\begin{eqnarray}
 \xi(T) \sim \exp(2J/T),
\end{eqnarray}
where $T$ is the temperature measured in a unit of $k_B = 1$. 
As the temperature is decreased further, the correlation along the transverse direction develops and the long-range order will take place at $T = T_c$ in the thermodynamic limit. 
However, if $N < \xi(T_c)$, the spins on each of the chains will order ferromagnetically even at $T > T_c$ and they will behave like a single-spin. So that 
the system will not show true critical behavior of the two dimensional model. 
Therefore, we must treat the lattice with $N \gg \xi(T_c) \gg 1$ and much CPU time is necessary to make the MC simulation. 
If, however, there exists some algorithm which leads an equilibrium spin configuration of individual chain, we may reduce the CPU time, 
because a large MC step is wasted to equilibrate the spin configuration of every chain. Here we propose that algorithm. 

We consider the two dimensional Ising model of eq.(1) on a rectangle lattice of $M \times N$. For simplicity, we impose the periodic boundary condition to the transverse direction, and the open boundary condition to the longitudinal direction. We consider the chain of $N$ spins along the longitudinal direction whose Hamiltonian is given by 
\begin{eqnarray}
 {\cal H}_j &=& -J \sum_{i=1}^{N-1} \sigma_i \sigma_{i+1} 
              -  \sum_{i=1}^N h_i \sigma_i,  \\
         h_i&=& J' (S_{ij+1} + S_{ij-1}),
\end{eqnarray}
where $\sigma_i \equiv S_{ij}$. Then the problem is to determine the spin configuration of the one dimensional chain with random fields {$h_i$}. This can be readily done as follows. 

We define the function:
\begin{eqnarray}
f_n(\sigma) &=& \sum_{\sigma_1,\sigma_2,\cdots,\sigma_{n-1}}\exp
          \{K(\sigsig{1}{2}+\sigsig{2}{3}+\cdots+\sigsig{n-1}{}) \nonumber \\
          &+&(C_1\sigma_1+C_2\sigma_2+\cdots+C_n\sigma)\}, 
\end{eqnarray}
where $ K = J/T$ and $C_n= h_n/T$.
$f_n(\sigma)$ can be readily obtained from the recursion formula 
\begin{eqnarray}
f_n(\sigma) = \sum_{\sigma_{n-1}} f_{n-1}(\sigma_{n-1}) 
         \exp \{ K\sigsig{n-1}{} + C_n\sigma \} \;\;\; n \geq 2, \nonumber \\
\end{eqnarray}
with  $f_1(\sigma)= \exp( C_1\sigma)$. 
Then the probabilities $P_N(\sigma_N = \pm 1)$ of the end spin being 
$\sigma_N = \pm1$ are given as 
\begin{eqnarray}
       P_N(\sigma_N = \pm 1) = \frac{f_N(\sigma_N = \pm 1)}{Z_N}, 
\end{eqnarray}
where $Z_N ( \equiv f_N(1) + f_N(-1))$ is the partition function of the chain. 
Then we can determine the value of $\sigma_N$ by using a uniform random number 
between 0 to 1. The next step is to determine the value of $\sigma_{N-1}$ under the condition that the value of $\sigma_N$ is fixed either $\sigma_N = 1$ or $\sigma_N = -1$. We can readily do it by changing the effective field on $\sigma_{N-1}$ from $h_{N-1}$ to $h_{N-1} + J\sigma_N$ and the function $f_{N-1}(\sigma)$  to 
\begin{eqnarray}
          \tilde{f}_{N-1}(\sigma) &=& \sum_{\sigma_{N-2}} f_{N-2}(\sigma_{N-2})                                                             \nonumber \\
        &\times& \exp \{ K\sigsig{N-2}{} + C_{N-1}\sigma + K \sigma \sigma_N \}                                                             \nonumber \\                  &=& f_{N-1}(\sigma) \exp(K\sigma \sigma_{N}).  
\end{eqnarray}
Then the probabilities $P_{N-1}(\sigma_{N-1} = \pm 1)$ are given as
\begin{eqnarray}
  P_{N-1}(\sigma_{N-1} = \pm 1) = \frac{\tilde{f}_{N-1}(\sigma_{N-1} = \pm 1)}                                         {\tilde{Z}_{N-1}},  
\end{eqnarray}
where
\begin{eqnarray}
      \tilde{Z}_{N-1} &=& \sum_{\sigma_{N-1}}\tilde{f}_{N-1}(\sigma_{N-1}) 
                                                                \nonumber \\
                       &=& f_{N}(\sigma_{N})\exp(-C_N \sigma_{N}), 
\end{eqnarray}
and the value of $\sigma_{N-1}$ can be determined. Repeating this procedure from $N-1$ to 1, we can determine the spin configuration of the chain. 
Using this algorithm, the spin configuration of the lattice is updated from chain to chain.

Note that the spin configuration of the chain $P(\sigma_1,\sigma_2,...,\sigma_N)$ realized in this procedure accords with the Boltzmann weight. That is, 
\begin{eqnarray}
       P(\sigma_1,\sigma_2,\cdot \cdot \cdot, \sigma_N) 
 &=& \frac{\tilde{f}_1(\sigma_1) } { \tilde{Z}_1 } 
   \frac{\tilde{f}_2(\sigma_2)}{\tilde{Z}_2} \cdot \cdot \cdot 
   \frac{f_N(\sigma_N)}{Z_N} \nonumber \\
 &=& \frac{1}{Z_N} \exp\{K\sum_{n=1}^{N-1}\sigma_n\sigma_{n+1} + \sum_{n=1}^NC_n\sigma_n\} \nonumber \\
 &=& \frac{1}{Z_N}\exp(-{\cal H}_j/T).
\end{eqnarray}
So it is natural to call this method  as {\it a chain heat bath method} or more generally {\it a cluster heat bath method} and to abbreviate it to a {\it CHB} method.
\begin{figure}
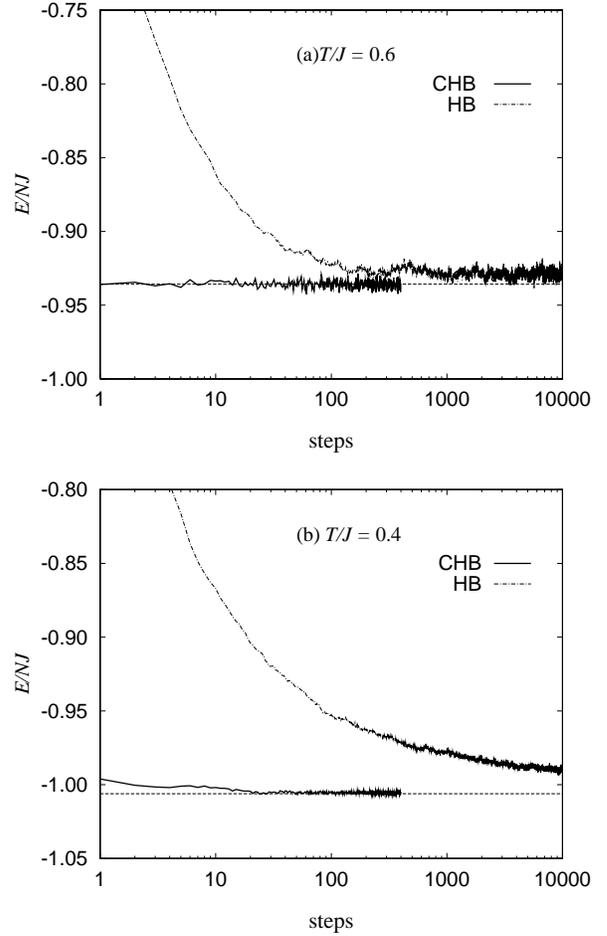

 \epsfxsize = 8cm
 \epsffile{relax0.6.epsi}

 \epsfxsize = 8cm
 \epsffile{relax0.4.epsi}
\caption{The relaxation processes in the lattice with $50 \times 500$ at (a)$T = 0.6J$ and (b)$T = 0.4J$. The broken line is the exact value by Onsager(ref.10).}
\label{fig:relax}
\end{figure}

\begin{figure}
  \begin{tabular}{ccc}
     (a) HB &  & \\
     \epsfxsize = 2cm
     \epsffile{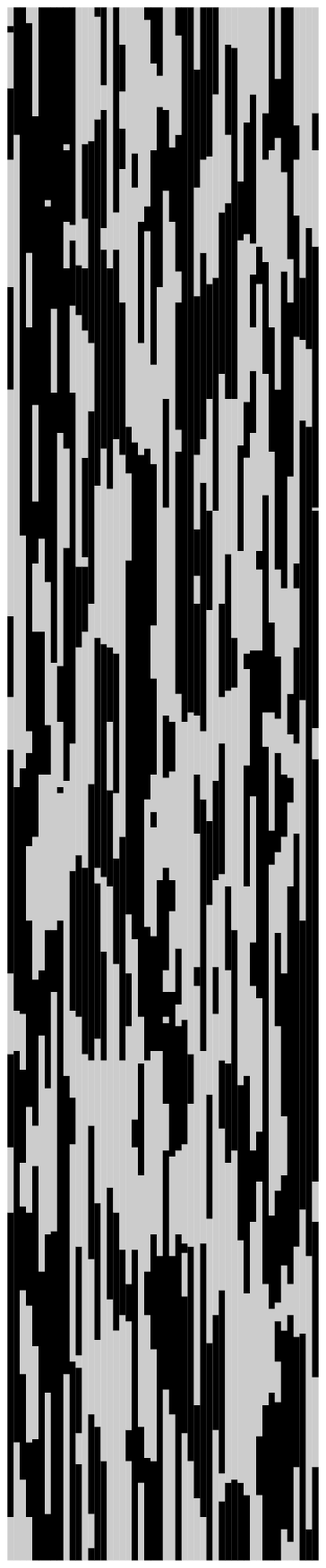} &
     \epsfxsize = 2cm
     \epsffile{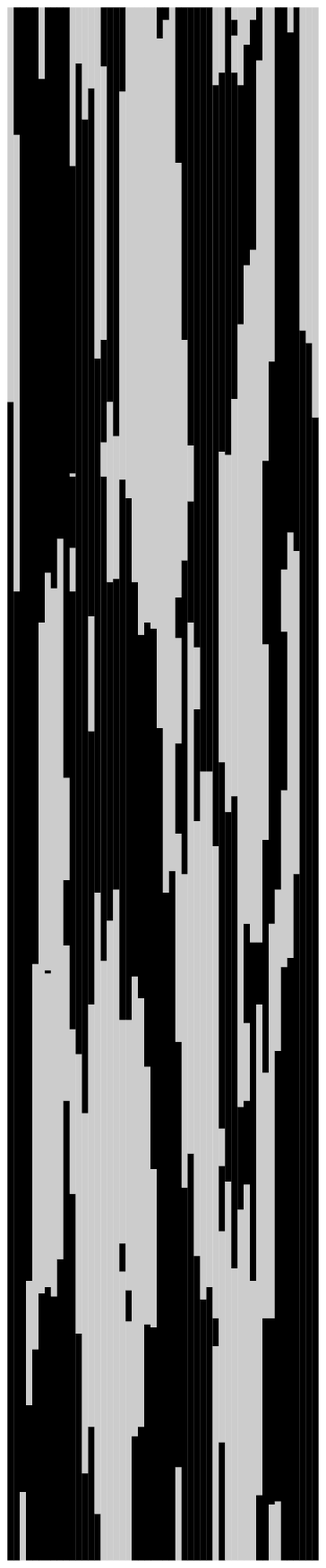} &
     \epsfxsize = 2cm
     \epsffile{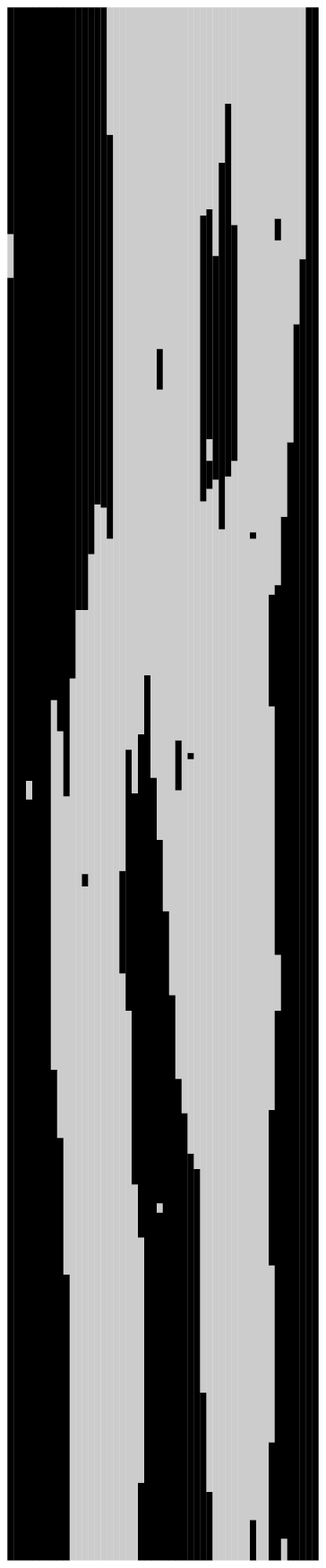} \\
     100 step & 1000 step & 10000 step\\
       & & \\
       & & \\
     (b) CHB &  & \\
     \epsfxsize = 2cm
     \epsffile{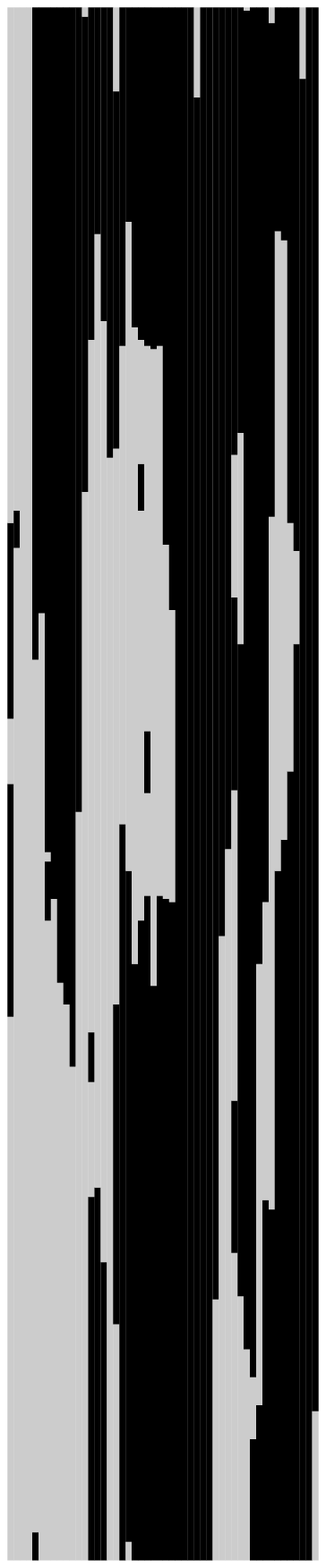} &
     \epsfxsize = 2cm
     \epsffile{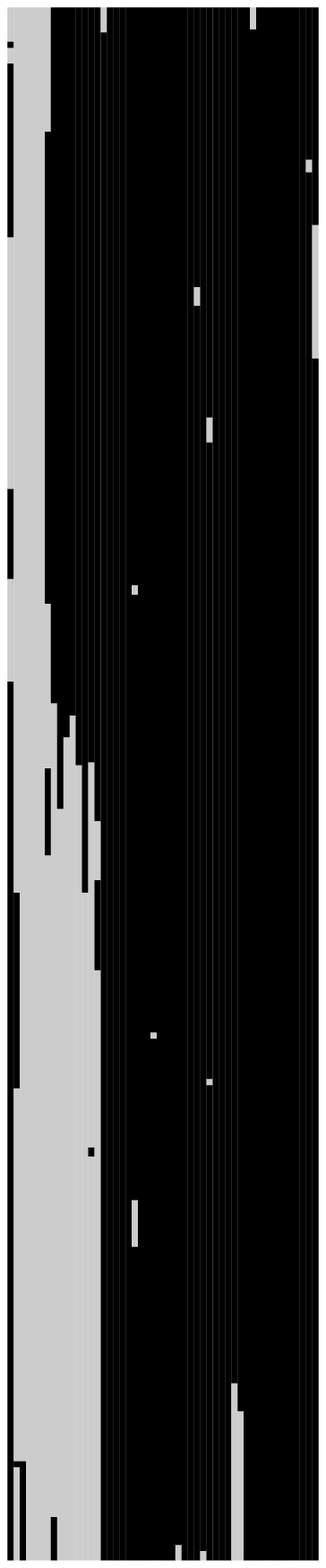} &
     \epsfxsize = 2cm
     \epsffile{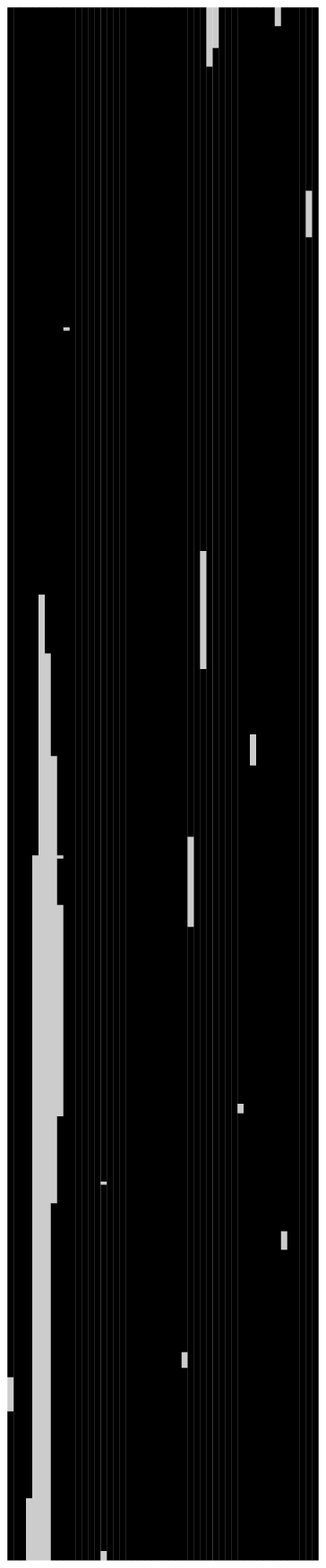} \\
      2 step & 8 step & 32 step
  \end{tabular}
\caption{Snapshots of the spin configuration in the lattice with $50 \times 500$ at $T=0.4J$, where the longitudinal lattice space is reduced to fifty percent. Up and down spins are described by black and gray points, respectively. }
\label{fig:snap}
\end{figure}

We apply the CHB method to the model (1) with $J'=0.01J$. The transition temperature is obtained from eq.(17) in ref.10 as $T_c \sim 0.51J$.
We examine the relaxation of the energy starting from a random spin configuration. In Figs. \ref{fig:relax}(a) and (b), we present typical results of the MC step dependence of the energy at temperatures above and below $T_c$, respectively, together with those by a conventional heat bath(HB) method. 
We find that the CHB method reproduces the exact result within a MC step much smaller than that of the HB method. 
We need only 1 or 2 MC steps to get the equilibrium value for $T > T_c$ and several dozens of MC steps for $T < T_c$, whereas we could not get it within 10000 MC steps in the HB method even for $T > T_c$. 
In Figs. \ref{fig:snap}(a) and (b), we present snapshots of the spin configuration for $T < T_c$ at different MC steps, $t$, in the HB and the CHB methods, respectively. In the CHB method, even at $t = 2$, every chain seems to reach its equilibrium state, and domains are formed. Smaller domains shrink and disappear at $t \sim 35$. On the other hand, in the HB method, even at $t = 1000$, many of the chains are still not in their equilibrium state. These facts clearly reveal that the fast relaxation in the CHB method comes from the instantaneous relaxation of every chain. 

\begin{figure}
\epsfxsize=8cm
\epsffile{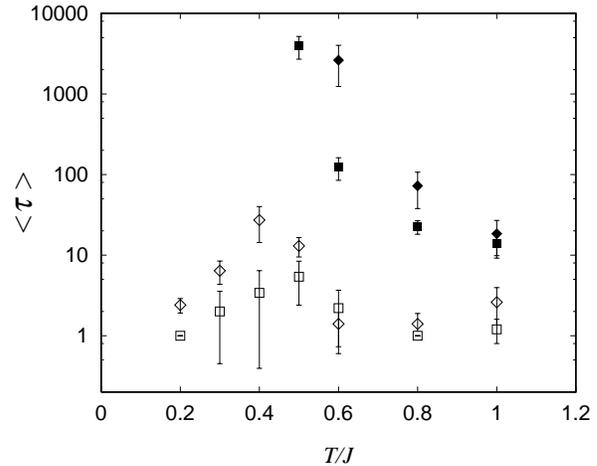}
\caption{Temperature dependences of the relaxation time  $\ave{\tau}$. The symbols of square and diamond mean those for lattices with $26 \times 250$ and $50 \times 500$, and those of filled and open ones mean those in the HB and the CHB methods, respectively.}
\label{fig:relax_all}
\end{figure}

We tentatively define the relaxation time $\tau$ as the step at which the 
energy firstly reaches the exact value. We made five runs starting from different random spin configurations and estimated it. In Fig. \ref{fig:relax_all}, we plot their average $\ave{\tau}$ and standard deviation $\itdelta\tau$ as functions of the temperature. 
In the HB method, $\ave{\tau}$ increases rapidly as the temperature is lowered towards $T_c$ and a large size dependence is seen. 
In the CHB method, for $T > T_c$, $\ave{\tau}$ depends little on both the temperature and the lattice size. It once increases at $T \sim T_c$ then decreases as the temperature is decreased further.  Even at $T \sim T_c$, it is less than several dozens of MC steps.

\begin{figure}
\epsfxsize=8cm
\epsffile{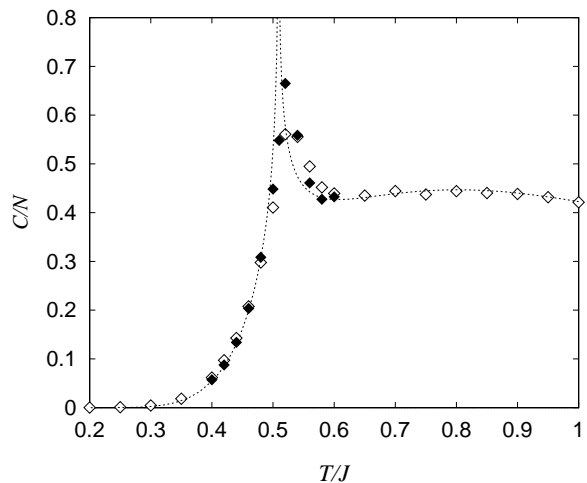}
\caption{The specific heat of the Ising model with $J' = 0.01J$ on the square 
lattice. Open and filled symbols mean those in the lattices with 
$50 \times 500$ and $100 \times 1000$, respectively, 
and a dotted line is the exact result by Onsager(ref.10).}
\label{fig:specific heat}
\end{figure} 

The specific heat calculated by the CHB method is shown in the Fig. \ref{fig:specific heat}. The result is good agreement with the exact value by Onsager in the whole temperature range. Especially, the sharp peak at $T = T_c$ is reproduced well.

We have proposed a cluster heat bath(CHB) method of a quasi-one dimensional Ising model and demonstrated that it reproduces exact results of the two dimensional Ising model within a MC step much smaller than that of the conventional MC method. The point of the method is to select one of the equilibrium spin configurations in each of the chains which are subjected by effective fields of surrounding spins. 
That is, each of the chains is always in its equilibrium state during the simulation.  Then the $d$ dimensional model becomes a $d-1$ dimensional model of fictitious spins with the freedom of $2^N$ and the simulation time is drastically reduced.  In fact, we could simulate a quasi-one dimensional Ising model on $36\times34\times1200$ hexagonal lattice which is a model of CsCoCl$_3$ and CsCoBr$_3$ within a reasonable computer time\cite{Koseki} and explain a change in the spin structure without any anomaly of the specific heat observed at a low temperature\cite{Yelon} as well as a usual phase transition at a higher temperature.

Finally, we should note that we could readily apply the CHB method to usual Ising models by treating columns of the Ising spins of $K \times L \times N (K, L = 1, 2, 3, 4,...)$ using a transfer matrix technique. The CHB method was found to be also effective for complex Ising models such as Ising spin glasses. Results will be reported elsewhere in a separate paper.\cite{Shira}  

\bigskip

We would like to thank Professor T. Shirakura and Dr. T. Nakamura for valuable discussions. This work was financed by a Grant-in-Aid for Scientific Research from the Ministry of Education, Science and Culture.

\end{multicols}

\end{document}